\documentclass[sigconf]{acmart}

\usepackage{graphicx}
\graphicspath{ {images/} }

\usepackage{amssymb}
\usepackage{algorithm}
\usepackage{algpseudocode}
\usepackage{algorithmicx}
\usepackage{amsmath}
\usepackage{multirow}

\begin{document}
\title{BinPro: A Tool for Binary Source Code Provenance} 


\renewcommand{\dbltopfraction}{1.0}
\renewcommand{\topfraction}{1.0}
\renewcommand{\bottomfraction}{1.0}
\renewcommand{\textfraction}{0.2}



\newcommand{\scbf}[1]{\vspace {4pt}\noindent{\textbf{#1.}}}

\newcommand{\myfig}[5]
{
\begin{figure}[t]
\begin{center}
\ifpdf
\includegraphics[width=#4\linewidth]{#1}
\else
\includegraphics[width=#4\linewidth]{#1}
\fi
\end{center}
\vspace{#5}
\caption{#2}\label{#3}
\vspace{-6pt}
\end{figure}
}

\newcommand{\myfigwide}[5]
{
\begin{figure*}[t]
\begin{center}
\ifpdf
\includegraphics[width=#4\linewidth]{#1}
\else
\includegraphics[width=#4\linewidth]{#1}
\fi
\end{center}
\vspace{#5}
\caption{#2}\label{#3}
\vspace{-6pt}
\end{figure*}
}

\newcommand{\mysubfigtwo}[8]
{
\begin{figure}
        \centering
        \begin{subfigure}[b]{0.25\textwidth}
                \centering
                \includegraphics[width=\textwidth]{#1}
                \caption{#2}
                \label{#3}
        \end{subfigure}%
        ~ 
        \begin{subfigure}[b]{0.25\textwidth}
                \centering
                \includegraphics[width=\textwidth]{#4}
                \caption{#5}
                \label{#6}
        \end{subfigure}
        ~ 
        \caption{#7}\label{#8}
        \vspace{-10pt}        
\end{figure}
}

\newcommand{\mysubfigthree}[9]
{
\def\tempa{#1}
\def\tempb{#2}
\def\tempc{#3}
\def\tempd{#4}
\def\tempe{#5}
\def\tempf{#6}
\def\tempg{#7}
\def\temph{#8}
\def\tempi{#9}
\mysubfigthreecont
}

\newcommand{\mysubfigthreecont}[2]
{
\begin{figure*}
        \centering
        \begin{subfigure}[b]{0.3\textwidth}
                \centering
                \includegraphics[width=\textwidth]{\tempa}
                \caption{\tempb}
                \label{\tempc}
        \end{subfigure}%
        ~ 
        \begin{subfigure}[b]{0.3\textwidth}
                \centering
                \includegraphics[width=\textwidth]{\tempd}
                \caption{\tempe}
                \label{\tempf}
        \end{subfigure}
        \begin{subfigure}[b]{0.3\textwidth}
                \centering
                \includegraphics[width=\textwidth]{\tempg}
                \caption{\temph}
                \label{\tempi}
        \end{subfigure}
        ~ 
        \caption{#1}\label{#2}
        \vspace{-5pt}
\end{figure*}
}

\newcommand{\mysubfigfourbox}[9]
{
\def\tempa{#1}
\def\tempb{#2}
\def\tempc{#3}
\def\tempd{#4}
\def\tempe{#5}
\def\tempf{#6}
\def\tempg{#7}
\def\temph{#8}
\def\tempi{#9}
\mysubfigfourboxcont
}

\newcommand{\mysubfigfourboxcont}[5]
{
\begin{figure*}
        \centering
        \begin{tabular}{cc}
        \begin{subfigure}[b]{0.5\textwidth}
                \centering
                {%
                \setlength{\fboxsep}{0pt}%
                \setlength{\fboxrule}{1pt}%
                \fbox{\includegraphics[width=\textwidth]{\tempa}}%
                }%
                \caption{\tempb}
                \label{\tempc}
        \end{subfigure}%
        &
        \begin{subfigure}[b]{0.5\textwidth}
                \centering
                {%
                \setlength{\fboxsep}{0pt}%
                \setlength{\fboxrule}{1pt}%
                \fbox{\includegraphics[width=\textwidth]{\tempd}}%
                }%
                \caption{\tempe}
                \label{\tempf}
        \end{subfigure}
        \\
        \begin{subfigure}[b]{0.5\textwidth}
                \centering
                {%
                \setlength{\fboxsep}{0pt}%
                \setlength{\fboxrule}{1pt}%
                \fbox{\includegraphics[width=\textwidth]{\tempg}}%
                }%
                \caption{\temph}
                \label{\tempi}
        \end{subfigure}
        &
        \begin{subfigure}[b]{0.5\textwidth}
                \centering
                {%
                \setlength{\fboxsep}{0pt}%
                \setlength{\fboxrule}{1pt}%
                \fbox{\includegraphics[width=\textwidth]{#1}}%
                }%
                \caption{#2}
                \label{#3}
        \end{subfigure}
        \end{tabular}
        \caption{#4}\label{#5}
        \vspace{-10pt}
\end{figure*}
}

\newcommand{\mysubfigsixbox}[9]
{
\def\tempa{#1}
\def\tempb{#2}
\def\tempc{#3}
\def\tempd{#4}
\def\tempe{#5}
\def\tempf{#6}
\def\tempg{#7}
\def\temph{#8}
\def\tempi{#9}
\mysubfigsixboxcont
}

\newcommand{\mysubfigsixboxcont}[9]
{
\def\tempj{#1}
\def\tempk{#2}
\def\templ{#3}
\def\tempm{#4}
\def\tempn{#5}
\def\tempo{#6}
\def\tempp{#7}
\def\tempq{#8}
\def\tempr{#9}
\mysubfigsixboxcontcont
}

\newcommand{\mysubfigsixboxcontcont}[2]
{
\begin{figure*}
        \centering
        \begin{tabular}{ccc}
        \begin{subfigure}[b]{0.32\textwidth}
                \centering
                {%
                \setlength{\fboxsep}{0pt}%
                \setlength{\fboxrule}{1pt}%
                \fbox{\includegraphics[width=\textwidth]{\tempa}}%
                }%
                \caption{\tempb}
                \label{\tempc}
        \end{subfigure}%
        &
        \begin{subfigure}[b]{0.32\textwidth}
                \centering
                {%
                \setlength{\fboxsep}{0pt}%
                \setlength{\fboxrule}{1pt}%
                \fbox{\includegraphics[width=\textwidth]{\tempd}}%
                }%
                \caption{\tempe}
                \label{\tempf}
        \end{subfigure}
        &
        \begin{subfigure}[b]{0.32\textwidth}
                \centering
                {%
                \setlength{\fboxsep}{0pt}%
                \setlength{\fboxrule}{1pt}%
                \fbox{\includegraphics[width=\textwidth]{\tempg}}%
                }%
                \caption{\temph}
                \label{\tempi}
        \end{subfigure}
        \\
        \begin{subfigure}[b]{0.32\textwidth}
                \centering
                {%
                \setlength{\fboxsep}{0pt}%
                \setlength{\fboxrule}{1pt}%
                \fbox{\includegraphics[width=\textwidth]{\tempj}}%
                }%
                \caption{\tempk}
                \label{\templ}
        \end{subfigure}
        &
        \begin{subfigure}[b]{0.32\textwidth}
                \centering
                {%
                \setlength{\fboxsep}{0pt}%
                \setlength{\fboxrule}{1pt}%
                \fbox{\includegraphics[width=\textwidth]{\tempm}}%
                }%
                \caption{\tempn}
                \label{\tempo}
        \end{subfigure}
        &
        \begin{subfigure}[b]{0.32\textwidth}
                \centering
                {%
                \setlength{\fboxsep}{0pt}%
                \setlength{\fboxrule}{1pt}%
                \fbox{\includegraphics[width=\textwidth]{\tempp}}%
                }%
                \caption{\tempq}
                \label{\tempr}
        \end{subfigure}
        \end{tabular}
        \vspace{-8pt}
        \caption{#1}\label{#2}
\end{figure*}
}

\newcommand{\BA}{{\em begin\_atomic}}
\newcommand{\EA}{{\em end\_atomic}}

\newcounter{claimcounter}[section]
\newtheorem{claim}{\sc Claim:}

\begin{abstract}
Enforcing open source licenses such as the GNU General Public License (GPL), analyzing a binary for possible vulnerabilities, and code maintenance are all situations where it is useful to be able to determine the source code provenance of a binary.  While previous work has either focused on computing binary-to-binary similarity or source-to-source similarity, BinPro is the first work we are aware of to tackle the problem of source-to-binary similarity.  BinPro can match binaries with their source code even without knowing which compiler was used to produce the binary, or what optimization level was used with the compiler.  To do this, BinPro utilizes machine learning to compute optimal code features for determining binary-to-source similarity and a static analysis pipeline to extract and compute similarity based on those features.  Our experiments show that on average BinPro computes a similarity of 81\% for matching binaries and source code of the same applications, and an average similarity of 25\% for binaries and source code of similar but different applications.  This shows that BinPro's similarity score is useful for determining if a binary was derived from a particular source code.
\end{abstract}

\author{
Dhaval Miyani, Zhen Huang, David Lie
}
\affiliation{University of Toronto}
\email{dhaval.miyani@mail.utoronto.ca, z.huang@mail.utoronto.ca, lie@eecg.toronto.edu}

\maketitle


\section{Introduction}

There are a number of situations where one wants to determine the source code provenance of a binary -- that is, the source code from which a binary originates.  This can help determine the authorship, intellectual property ownership rights and trustworthiness of a binary.  For example, the GNU General Public License (GPL)~\cite{gnu} requires any one who modifies and distributes an application protected by the license to make the source code modifications publicly available -- a requirement that is often violated, as demonstrated by numerous GPL infringement lawsuits~\cite{oneill_what_2010,versata,busybox}.  To determine if a distributed binary infringes, a plaintiff must show that the binary was derived from the source code of an application protected by the GPL, often without the cooperation of the producer of the binary.  Since GPL only requires source code to be shared if the binaries are released to the public, source code provenance can also be used by organizations and developers as an additional check to ensure that they do not mistakenly release binaries derived from GPL source code.

In other instances, a user or an organization may wish to know the source code provenance of a binary for security or maintenance purposes.  Knowing the identity of the source code of a binary may help identify possible vulnerabilities.  For maintenance purposes, it can help identify possible ways of updating and maintaining the binary if the original source code for the binary has been lost.  

In an ideal world, the translation from source code to binary is a straightforward process and determining whether a binary was derived from a particular source code is simply a measurement of the similarity of the source code and binary.  For example, one might compute a call-graph of both binary and source code and compute the similarity of the graphs using one of a number of well-known methods~\cite{ullmann1976algorithm,hungarian-algorithm}.  However, in reality, optimization applied during the compilation process can result in a binary that has a significantly different call-graph and subroutines than the original source code.  These optimizations have traditionally been the bane of systems that attempt to compute binary-to-binary similarity~\cite{BinDiff-paper,BinSlayer,BinHunt,blex}. Unfortunately, the heuristics compilers use to determine when and where optimizations are applied are opaque and difficult to predict, often being sensitive to a large number of factors in the source code.

Since determining the source code provenance of a binary requires access to the source code features during the matching process, we expect that the accuracy of matching in this case to be better than binary-to-binary matching.  In this work, we show this is indeed the case, especially in the case when code optimizations such as function inlining are applied.  Our most surprising result is that determining when and where a compiler will apply inlining can be done {\em without} analysis of the compiler code at all.  Instead, we generate a training set of optimized and unoptimized binaries by simply compiling unrelated applications, and use this to train a machine learning model, which can then be used to predict when optimization will be applied by the compiler.  As one would expect, this works fairly well when the compiler used for training the model is the same as the one used to produce the binary, but surprisingly, it works fairly well even when it is a different compiler, or the same compiler used with different optimization levels.

We demonstrate the utility of this approach with a tool called BinPro, which, given a binary and a source code, computes a similarity score for them.  A high similarity score indicates that the binary is very likely to have been the result of compiling the source code, while a low similarity score indicates that the binary was likely compiled from some other source code.  We evaluate BinPro on a corpus of applications and libraries and demonstrate that BinPro's similarity score correlates well with whether binaries and source code really match even across different compilers and optimization levels. 

\vspace{2pt}
\noindent In summary, we make the following contributions:
\begin{enumerate}
	\item We present BinPro, which is the first technique we are aware that is able to match a program's source code and binary using a novel combination of machine learning and static analysis.
	\item We evaluate BinPro on a corpus of 10 executable applications and 8 libraries, and demonstrate that BinPro produces similarity scores ranging from 59\% to 96\% and an average of 81\% for matching binaries and source code, and scores ranging from 10\% to 43\% with an average of 25\% for non-matching binary and source code.  This demonstrates that BinPro is able to determine whether a binary was derived from a particular source code or not.
	\item We show that BinPro is insensitive to different compilers or compiler optimization levels used to compile the binary.  We further show that BinPro's machine learning models, when trained on one compiler, GCC, is able to predict when function inlining optimizations will be applied by a completely different compiler, the ICC Intel compiler.  
\end{enumerate}

The remainder of the paper is organized as follows. First, we review related work on computing binary-to-binary similarity and source-to-soruce similarity in Section~\ref{sec:relatedWork}.  Then, we describe the design of BinPro Section~\ref{sec:design}.  Implementation details are given in Section~\ref{sec:implementation} and we evaluate BinPro's effectiveness in Section~\ref{sec:evaluation}.  Finally, we conclude in Section~\ref{sec:conclusion}.

\section{Related Work}
\label{sec:relatedWork}

Determining whether two programs are equivalent or not is reducible to the halting problem, and is thus undecidable.  For the same reason, it is difficult to prove that compilers produce binary code that is equivalent to the input source code~\cite{necula_translation_2000,leroy_formal_2009}.  However, despite these difficulties, researchers have still made significant progress with a number of proposals for practically measuring the similarity of two programs. Previous work in computing program similarity breaks down into three major sub-problems: a) measuring the similarity of sections of binary with corpus of source code, b) measuring the similarity of two binaries and c) measuring the similarity of two sections of source code.  We review related work in the other two subproblems below.

\subsection{Measuring binary$-$source code similarity}
Hemel et al., developed the Binary Analysis Toolkit (BAT), a system for code clone detection in binaries to detect GPL violations~\cite{hemel2011finding}. It recursively extracts strings from a binary, such as a firmware image. It attempts to detect cloning of code by matching strings with a database of packages of GPL projects. It also attempts to detect similarity through data compression and binary delta techniques. While BAT solves slightly different and harder problem in that it determines whether a binary code contains statically linked library code, it has high false negatives. This means that, unlike BinPro, BAT finds an incorrectly matching source code for a given binary.  BinSourcerer~\cite{BinSourcerer} and RESource~\cite{rahimian2012resource} identify individual source functions in a binary, rather than computing similarity between the entire binary and source code. However, there are no measurements on real applications on how well their approaches performs.

Di Penta et al.~\cite{di2010identifying} describe a tool that identifies licensing of JAVA archives (JARs) by analyzing JAVA .class files and submitting detected package and class names to Google Code Search to identify the provenance of the code. However, it is much easier to decompile JAVA byte-code and generate code almost similar to the original, rather than C/C++ machine code, which is in the case of BinPro. Davis et al.~\cite{davies2011software} also provides similar approach for finding provenance of JAVA applications.

\subsection{Measuring binary similarity}

Much of the work on measuring binary code similarity has the goal of classifying malware into families by similarity, or to detect if two binaries may be equivalent for vulnerability detection.  As a result, the goal is to detect semantic similarity between binaries even if the actual instructions or instruction sequences are different.  The work in this area can be broadly classified into static and dynamic approaches

\scbf{Static approaches} Static approaches extract features from binaries, such as control flow graphs (CFG) and then compare the graphs of the two binaries.  
Zynamics BinDiff~\cite{BinDiff-paper} is an industry standard state-of-the-art binary diff-ing tool. BinDiff matches a pair of binaries using a variant of graph-isomorphism algorithm. BinDiff extracts CFGs from two binaries and tries to match functions between each binary using heuristics. The major drawback of BinDiff is that it performs extremely poorly when comparing two binaries that are compiled with different optimization levels or with different compilers.  Even though the programs may be functionally equivalent, many compiler optimizations affect program CFGs greatly and thus make graph matching ineffective.  

Inspired by BinDiff,  BinSlayer~\cite{BinSlayer} and Pewny et al~\cite{pewny_crossarchitecture_2015} perform bipartite matching using the Hungarian algorithm.  This allows them to be more resilient to CFG changes due to local compiler optimizations.  DiscovRE~\cite{eschweilersebastian_discovre_2016} uses an even looser matching algorithm to match binaries using structural and numeric features of the CFG.  However, despite the increased accuracy despite CFG transformations, neither approach handles function inlining very well, which introduces code from a inlined callee function into the caller function's CFG.

While BinPro does rely on function-level code features, BinPro does not use features from a function's CFG for matching.  Our experiments empirically show that CFG features are unreliable for program matching, and thus BinPro excludes them from its matching strategy and relies on other features instead.  In addition, BinPro uses machine learning to predict when functions might be inlined by the compiler, allowing BinPro to properly compute similarity even for inlined functions.

\scbf{Dynamic approaches} Rather than perform a similarity measurement over statically extracted graphs, these methods measure similarity by comparing the execution of programs.  Egele et al. propose Blanket Execution, BLEX~\cite{blex}, an engine to match functions in binaries, with the goal of either classifying malware or aiding automatic exploit generation. BLEX uses dynamic equivalence testing, which executes the code of both binaries and compares the effects of the executions to determine the similarity of the binaries.  BLEX attempts to ensure that every basic block is executed at least once per function.  However, for large functions, BLEX may not complete and in practice, analysis is halted when the execution time exceeds a specified timeout value or a maximum of 10,000 instructions have been executed.  

BinHunt~\cite{BinHunt} generalizes individual executions by using symbolic execution and a theorem prover to determine semantically equivalent basic blocks. However, BinHunt suffers from performance bottlenecks due to its symbolic execution engine and it is unclear whether it can scale to large, real-world applications.  In general, while execution enables dynamic approaches to sidestep the unpredictability of compiler optimizations, they often come with higher costs in terms of execution time and required compute resources for large programs.  In addition, reliable measurement requires the execution of a reasonable fraction of the paths in a program, but the number of program paths tends to grow exponentially with program size.  

BinPro's analysis uses only statically extracted code features and does not consider individual instructions, making its analysis much closer to the static approaches outlined above.  As a result, BinPro easily scales to program sizes in the hundred's of thousands of lines of code with only modest resource requirements (10's of gigabytes of RAM).  Despite this, BinPro's innovations allow it to be resilient to compiler optimizations.  

A strawman alternative to BinPro might be to compile the source code to binary and then perform one of the previously proposed binary-to-binary comparisons.  However, because BinPro has access to features in source code, we show that these features can be used to better predict compiler optimizations such as inlining, which have frustrated previous static approaches.  

\subsection{Measuring source similarity}

The motivations for measuring source code similarity are usually to detect bugs that result from copying code or to detect code plagiarism. CCFinder~\cite{kamiya2002ccfinder} and CP-Miner~\cite{li2004cp} analyze the token sequence produced by a lexer and deduce equivalence when duplicate token sequences, which indicate potential code clones.  Such systems however, can only detect exact copies, otherwise known as code clones.

More sophisticated systems use abstractions of the source code to gain resilience to syntactic differences in source code that don't affect the semantic meaning of the code.  For example, ReDeBug~\cite{jang2012redebug} normalizes tokens to remove semantically irrelevant information, such as whitespace, comments and variable names.  It then uses normalized token sequences to find code clones with the goal of scalably detecting unpatched code clones in OS-distribution scale code bases.  DECKARD~\cite{jiang2007deckard} uses abstract syntax trees as an abstraction for computing source code similarity.  The trees are represented as numerical vectors, which are clustered using Euclidean distance. Yamaguchi et al.~\cite{yamaguchi2012generalized} extend this idea by determining structural patterns in abstract syntax trees, such that each function in the code could be described as a mixture of these patterns. This representation enables identifying code similar to a known vulnerability by finding functions with a similar mixture of structural patterns.  More recently, VulPecker applies machine learning to the problem of source code similarity for detecting vulnerabilities~\cite{li_vulpecker:_2016}.

While related, these systems solve a different problem than BinPro in that they look for semantic similarity in two code samples written in the same language.  As a result, they can rely on simpler syntactic comparisons without having to perform any deep semantic analysis.  In contrast, BinPro seeks to detect similarity between source code and binary, which are syntactically different.  As a result, BinPro uses semantic code features, such as string and integer constants, function and library calls and function declaration information to compute similarity.

\section{Design} 
\label{sec:design}

\subsection{Overview} 
\label{sec:overview}

\myfigwide{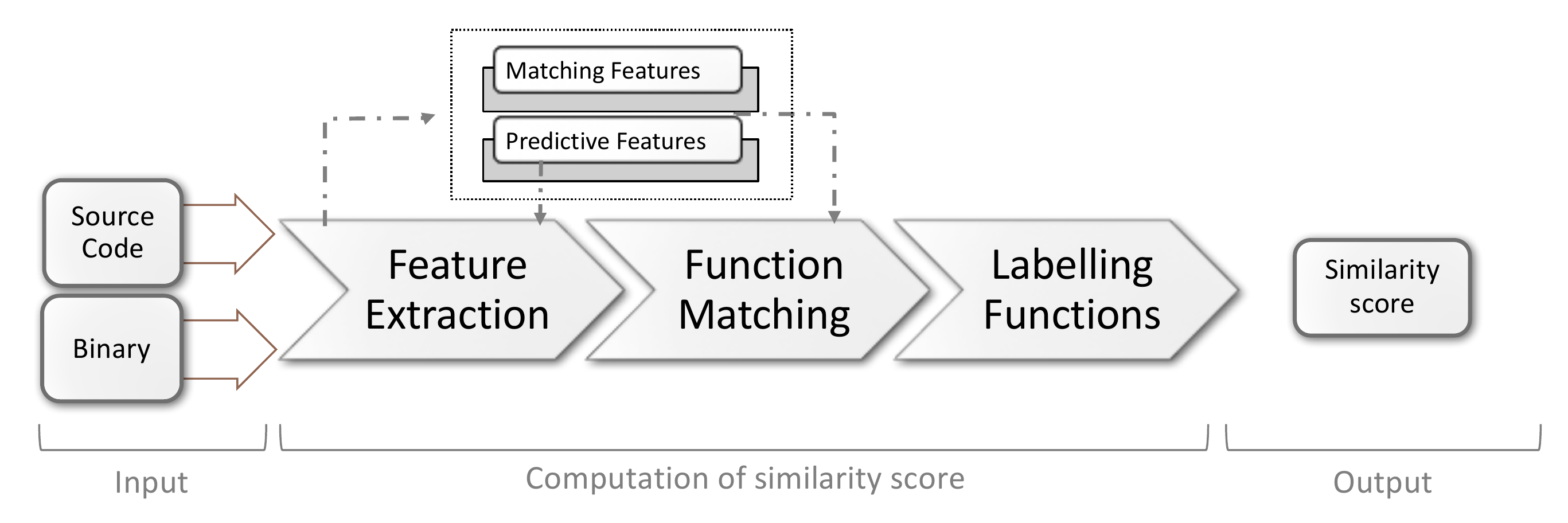}{High-level work flow of BinPro.}{fig:workFlow}{0.93}{-20pt}

BinPro takes as input a binary and source code, and computes a similarity score that is the percentage of binary code functions that match source code functions.  The higher the similarity score, the more likely that the binary was compiled from the given source code.  BinPro is agnostic to superficial modifications, such as changing or removing non-compiled sections of source code (i.e. comments or white space) or even changing variable names, function names or the order of declarations.  However, more significant changes, such as changing program structure, constant values or strings will affect the similarity score computed by BinPro.  In most cases, we expect that the goal is to determine as closely as possible, whether the given source code was used to compile the binary, as this is what is usually needed for code maintenance or to detect GPL violations.  

We make several assumptions in this work. We assume that binaries are stripped of symbols and have been compiled with a compiler that the customer does not have access to, which might arbitrarily apply a variety of compiler optimizations during the process of transforming source code into a binary. As a result, the major challenge for BinPro is differentiating between differences in binaries and source code that are due to legitimate compiler optimizations. To overcome this, BinPro identifies code features that are invariant under most compiler optimizations.  For the remaining optimizations that do alter these features, we use machine learning to train BinPro to predict when these optimizations are likely to be applied to allow BinPro to account for them. We further assume that the binary is not obfuscated that it can be reliably disassembled using a tool such as IDAPro~\cite{ida}.

Figure~\ref{fig:workFlow} shows the high-level work-flow of BinPro.  First, a set of code features are extracted from both binary and source code.  Then BinPro computes an optimal bipartite match between binary and source code functions using those features.  Finally, based on their similarity, each pair of matched binary and source code functions is labeled as uniquely matched, multi-matched (where a binary matches several source code functions equally well) or unmatched (where a binary code function doesn't match any source code function).  The percentage of uniquely matched functions over the total number of functions in the binary is the similarity score for the binary and source code.  We now describe each of these stages in more detail.

\subsection{Feature extraction} 
\label{sec:featuresExtraction}

To compute an optimal bipartite match, BinPro must compute a weight that describes the similarity of each pair of binary and source code functions.  To do this, BinPro scans binary and source code to obtain a list of functions in each.  Each source code function is uniquely identified by its name and each binary code function by its starting address.  
The code associated with each function is not recorded since binary code cannot be directly compared with source code for similarity.  Instead, BinPro abstracts the binary or source code into a common set of {\em matching features} for each binary and source code function. The weights for each function-pair can then be computed by comparing their respective matching features.

Bipartite matching implicitly assumes that there is a one-to-one correspondence between binary and source code functions -- an assumption that is broken by function inlining optimization.  To address function inlining, BinPro uses a different set of {\em predictive features} in the source code to predict which function calls a compiler is likely to inline.  Based on these predictions, BinPro then creates {\em pseudo-inlined functions} that combine the matching features of the inlined callee function and the parent caller function, and adds them to the list of source code functions.  The psuedo-inlined functions do not actually have any source code, but exist in the list of functions as a candidate source function that BinPro can match a binary function with during the function matching stage.  If the function inlining occurred during compilation, the resulting inlined binary code function will match the added inlined function.

We summarize the code features that BinPro uses in Table~\ref{table:featureSet}. 
Because compilers decide whether to inline functions based on the source code of the functions, these predictive features are extracted only from source code.  In contrast, matching features are extracted from both source code and binaries.  We now describe in detail how each of the features is extracted.

\scbf{Matching features} Features are extracted from binaries using a binary analysis tool and from source code by a compiler tool (Details on the tools are given in Section~\ref{sec:implementation}).  Source code features are extracted after pre-processing.  Constants string literals can be extracted from the source code directly.  In binaries, the addresses of constant string literals can be found in the constants section of the binary and use-def analysis can be applied in both binaries and source code to find all uses of string literals in functions.  Integer constants used in functions are extracted directly from program statements in the source and assembly instructions in the binary.  BinPro ignores common constants 0, 1 and -1 and does not include those in the extracted set of features.  To extract the number of function call arguments from binaries, BinPro looks for definitions of the registers used to pass arguments to callees before a function call to determine the number of arguments passed in a function call, and uses those registers to determine the number of arguments used.  Compilers may optimize some of these use and definitions in certain cases and will cause BinPro to extract the wrong number of arguments for binary code functions.  The effect of this error is mitigated by the weight assigned to this feature in the machine learning phase described in Section~\ref{sec:machineLearningImplementation}.  Previous work ~\cite{Caballero_EECS} can be used to accurately extract the number of arguments feature from binaries.  The function call graph (FCG) features is a set of callers and callees for each function, represented as a set of function names in the source code, and a set of instruction addresses from the binary.  

Library calls can be identified and extracted as they are calls to functions outside of the binary or source code.  In practice, some compilers perform system-specific library call substitutions (i.e. replacing call to {\tt printf} in the source code with a call to {\tt puts} or {\tt vsprintf} in the binary).  BinPro is configured to ignore library calls that are substituted in this way.  Compilers also occasionally insert functions into the binary that did not come from the source code such as \texttt{\_\_stack\_chk\_fail}, which GCC inserts to detect stack overflow.  For our evaluation, BinPro is configured to ignore an empirically-determined set of 19 such functions.  

Compilers may also insert string literals into the binary that are not present in the source code, such as pre-processor defined strings, i.e. {\tt \_\_LINE\_\_} or {\tt \_\_FILE\_\_}.  Another more obscure case is that format strings can be substituted at compilation time when the resulting string could be statically resolved (i.e. {\tt sprintf(str, "\%d", 5)}).  However, these string substitutions are difficult to whitelist as the inserted string is dependent on both the source code and the environment where the binary was compiled (i.e. the compilation date or the compiler version).  As a result, BinPro's results are negatively affected by these literals and may lead to matching errors, but overall BinPro is able to accurately determine provenance despite errors introduced by this artifact.

A feature often used by other code similarity work is the control flow graph (CFG) of each function~\cite{BinDiff-paper,BinSlayer,pewny_crossarchitecture_2015}.  However, we find that due to the large number of local optimizations compilers perform, there is little correlation between source code and binary CFGs as we will elaborate further in Section~\ref{sec:machineLearningImplementation}.  As a result, BinPro does not rely on CFG features for computing similarity.


	\begin{table} [tb]
		\caption{Binary and source code features used by BinPro.}
		\begin{center}			
			\begin{tabular}{|l|p{0.6in}|p{0.6in}|}
				\hline
				\multicolumn{1}{|c|}{\textbf{Features}} &
				\multicolumn{1}{|p{0.6in}|}{\textbf{Matching}} & 
				\multicolumn{1}{|p{0.6in}|}{\textbf{Predictive}} \\ \hline
				String constants &  \multicolumn{1}{|c|}{\checkmark}  & \\ \hline
				Integer constants &  \multicolumn{1}{|c|}{\checkmark}  & \\ \hline
				Library calls &  \multicolumn{1}{|c|}{\checkmark}  &  \\ \hline
				FCG callers &  \multicolumn{1}{|c|}{\checkmark}  & \\ \hline
				FCG callees &  \multicolumn{1}{|c|}{\checkmark}  & \\ \hline
				\# of function args & \multicolumn{1}{|c|}{\checkmark}  & \\ \hline
				Static func & & \multicolumn{1}{|c|}{\checkmark} \\ \hline
				Extern func & & \multicolumn{1}{|c|}{\checkmark}  \\ \hline
				Virtual func & & \multicolumn{1}{|c|}{\checkmark}  \\ \hline
				Nested func & & \multicolumn{1}{|c|}{\checkmark}  \\ \hline
				Variadic args & & \multicolumn{1}{|c|}{\checkmark}  \\ \hline
				Recursion & & \multicolumn{1}{|c|}{\checkmark}  \\ \hline
				
			\end{tabular}
		\end{center}
		\label{table:featureSet}
		\vspace{-20pt}
	\end{table}

\scbf{Predictive features} Because BinPro has access to source code, it is able to use the source code features listed in Table~\ref{table:featureSet} to predict which functions are likely to be inlined.  Access to source code is one of the reasons why BinPro can achieve much better results than code-similarity tools that only work with binaries~\cite{BinDiff-paper,BinSlayer,BinHunt}.  


Compilers use these features to determine which functions will be inlined.  However, different compilers may make that determination differently, and even the same compiler may inline different functions depending on the level of optimization specified during compilation.  Since BinPro does not assume access to the compiler or environment used to build the binary, we cannot assume accurate information about which functions are inlined.  Instead, BinPro makes an approximate prediction by training a classifier on a generic compiler used to compile a set of benchmark applications.  Neither the applications nor the compiler need to be the same as the application and compiler on which BinPro is ultimately be used to perform source code provenance of the binary application.  The intuition is that all compilers follow similar principles of when to inline, which are applied independently of the application being compiled.  

We build a corpus of inlined and non-inlined functions extracted from a variety of applications and then use this set of applications to train an Alternating Decision Tree (ADTree) classifier.  We used a decision tree algorithm because we believe that it mirrors the logic that compilers tend to use to decide whether to inline a function or not.  To verify this hypothesis, we evaluated different machine learning algorithms from the Weka toolkit~\cite{weka} and found that ADTree was indeed the best classifier for this purpose.  The trained classifier is then applied to the source code and a list of functions that are likely to be inlined is identified.  As described above, additional pseudo-inlined functions are created and added to the list of source code functions.  We note that BinPro does not remove the original functions, which are predicted to be inlined, from the set of source code functions. Instead, BinPro keeps both versions of functions in the set -- pseudo-inlined functions and its corresponding original functions. In this way if a binary code function is not inlined then it will be matched with its original source code function while if it is inlined, then the resultant binary code function will be matched with the pseudo-inlined function.   
	
As a result, it is important that inlining prediction is accurate.  Under-predicting inlining will cause inlined binary code functions to not match to any source code function even if the source code and binary do match in reality.  Over-predicting inlining will cause a multitude of extra functions to be added to the source and increase the chances that a binary code function will falsely match one of these functions even if the binary and source code to not match in reality.

\subsection{Function matching}
\label{sec:matching}
\label{sec:matchingFCG}

BinPro performs function matching by computing the pairwise weight of each binary-source function pair and then performing an optimal bipartite match. By representing the similarity of the matching features between binary and source code functions as weights, BinPro can use a bipartite matching algorithm to find a pairing between binary and source code functions that minimizes the differences between the pairs.   Representing function matching this way lets BinPro use known bipartite matching solutions, such as the Hungarian algorithm~\cite{hungarian-algorithm}, which BinPro uses to find a sub-optimal solution in polynomial time, O(n\textsuperscript{3}).  

Since BinPro extracts FCGs for both binary and source, a reasonable alternative to bipartite matching might have been to start at the entry points of both binaries (i.e., {\tt main()}), and then perform a traversal of the call graph, checking each function as it is found~\cite{BinDiff-paper}.  However, this alternative will not work for three reasons.  First, a function could have several callees, and there must be a way to disambiguate them.  Second, edges in the call graph are often incomplete due to the use of computed function pointers, whose targets cannot be determined statically.  Third, source code equivalence for shared (or dynamically linked) libraries cannot use this alternative because libraries do not necessarily have a single entry point for graph traversal.  In summary, the graphs extracted between source and binary for entire programs pose problems that make graph-traversal based similarity impractical.  Bipartite matching uses features that include properties other than graph properties, that allows it to bootstrap the matching.  

This bootstrapping property is required because FCG features are not initially complete at the end of feature abstraction.  Because instruction addresses are used to represent the callees and callers of binary code functions, while functions names are used for source code functions, unless there is a mapping between instruction addresses and function names, there is no way to compare the caller and callee features for similarity.  To overcome this challenge, BinPro performs function matching iteratively.  Initially, FCG callers and callees are not used and BinPro relies on other features to compute the weights and perform bipartite matching.  Once some functions are matched, BinPro is then able to infer a mapping between function names and instruction addresses, which BinPro uses to update the caller and callee features and recompute weights.  Bipartite matching is then repeated iteratively until the matching reaches a steady state.  This entire process for function matching is illustrated algorithmically in Algorithm~\ref{algo:matchFunction}. 

\algrenewcommand{\algorithmiccomment}[1]{$\rhd$ #1}
\begin{algorithm} [tb]
	\caption{Function matching}
	\label{algo:matchFunction}
	\begin{algorithmic}
		\Function{FuncMatch\textsl{}}{$BinGraph$, $SrcGraph$}
		\Repeat
		\State $reGenerateCallerAndCalleeNodes(BinGraph)$
		\State $reGenerateCallerAndCalleeNodes(SrcGraph)$
		\State \Comment{Create weighted bipartite graph}
		\State $weights\gets \emptyset$
		\State $pairs \gets \emptyset$
		\For{$b \in BinGraph$}
		\For{$s \in SrcGraph$}
		\State $weights \gets compWeights(b, s)$
		\EndFor
		\EndFor
		\State \Comment{Run Hungarian and assign labels}
		\State $pairs \gets HungarianAlgorithm(weights)$
		\State $assignLabels(binGraph, pairs)$ 
		\Until{$steady\_state=true$}
		\EndFunction
	\end{algorithmic}
\end{algorithm}

Weights are calculated for each pair of binary and source code functions as follows:
\begin{equation}
	\label{eq:totalCost}
	\sum_{i=1}^{N} {w_i C_{f_i}}
\end{equation}
Where \(N\) is the total number of features, \(w\) is the weight coefficient for a feature and \(C_f\) is the cost for a feature. Cost for a feature of the two sets is the same as distance between them.

\(C_f\) is a value between 0 and 1 where smaller values indicate that the pair of functions is more similar. In the special case where the costs for all features is 1, we assign the weight for the pair to be a special amount \textit{MAX\_WEIGHT}, which indicates that the pair have nothing in common.  \textit{MAX\_WEIGHT} is chosen so that it is larger than the largest possible weight that can be computed from equation~\ref{eq:totalCost}, allowing BinPro to separate pairs that have no features in common from those with high cost during labeling.  

The way a cost for a feature is computed depends on whether the feature is a set (such as for string constants, integer constants, library calls, or FCG callers and callees) or whether it is a scalar value (such as for the number of function arguments).  For set features, BinPro uses a modified Jaccard to compute their weight.  Recall the standard Jaccard index:
\begin{equation}
	\label{eq:jaccard}
	 J(B, S) = \frac{| B \cap S |}{| B \cup S |} 
\end{equation}

Where \(B\) and \(S\) represent a set feature from a node in the binary graph and source graph, respectively. The problem with the standard Jaccard is that it computes the same cost regardless of whether there are extra elements in $B$ or  $S$.  However, compiler optimizations tend to remove code features instead of add them, so that a binary code function is more likely to have fewer features, and less likely to have extra features, than its corresponding source code function.  In addition, compilers are likely to only be able to remove a few features during optimization since they have to preserve the functionality of the original code.  

To take these effects into account, we modify the Jaccard index, as shown in equation~\ref{eq:costFeature}, so that when $B$ is a subset of $S$, then the cost is based on the elements in $S$ that would have to be removed to make $B$, and when $B$ is not a subset of $S$, the cost is based on the features found in $B$ that are not in $S$. 
\begin{equation}
 	\label{eq:costFeature}
 	C_f(B, S) = 
 	\begin{cases}
 		\frac{| S \cap B^{C} |}{| S |}	& \text{if } B \subset S \\\\
 		\frac{| B \cap S^{C} |}{| B |} & \text{otherwise}
 	\end{cases}
\end{equation}
Where $B^{C}$ and $S^{C}$ represent the set complement of $B$ and $S$ respectively.
 
For the number of function arguments, which is the only scalar feature, we assign a cost of 1 if the source code and binary code functions do not have the same number of arguments and a 0 if they do.  We treat variadic source code functions as if they have 6 arguments due to the number of registers used for passing arguments in x86-64, which is further discussed in section~\ref{sec:implementation-extracting}.


To compute the weight coefficients, we use Sequential Minimal Optimization (SMO)~\cite{smo-ml} to train a Support Vector Machine (SVM) \cite{svm-ml} machine learning classifier. Once the training is completed, we obtain the weight coefficient for each feature according to their importance as determined by the SVM. We discuss details on this training process in Section~\ref{sec:machineLearningImplementation}.


\subsection{Labeling functions} 
\label{sec:labeling}


Once weights are assigned to each edge in the bipartite graph, the Hungarian algorithm will determine an assignment that minimizes the total weights of all assigned function-pairs.  From this, one might assume that examining the weights of the function-pairs will indicate how well similar the binary and source code are overall.  However, on closer inspection we find that comparing weights across different binary-source code pairs is not particularly meaningful. Using the applications evaluated in Section~\ref{sec:evaluation}, we find that the average weight ranges from 0.6-3.8 with an average of 1.3 when the binary is compared with the source code from which it was compiled, while the average weight ranges from 2.2-3.9 with an average of 3.0 when the binary differs from the source code it is compared with.  While, as expected, the average weight for the case where binary and source code differ is greater, the overlap in the ranges of average weight between these two cases motivated us to use the percentage of matched functions in the binary as a measure of similarity instead.  

To compute this percentage, BinPro first labels each function-pair as matched, multi-matched or unmatched as follows.  If the weight of the edge between a binary code function and a source code function is unique among all other edges to that binary code function then it is labeled as matched.  If it is not unique -- i.e. the binary code function matches several source code functions equally well -- then the binary code function is labeled as multi-matched.  If the weight of the edge is  \textit{MAX\_WEIGHT},  it means that there are no common matching features between the pair of functions. Such functions are labeled as unmatched.  

Matched functions provide a mapping between the instruction address of binary code functions and the name of source code functions.  As mentioned earlier, this mapping is then used to recompute the cost of the callee and caller set features, which in turn allows BinPro to update the weights between binary and source code functions and continue to iteratively match functions.  

Finally, while functions are labels as unmatched and multi-matched, only the percentage of matched functions are used to compute the similarity score.

\section{Implementation}
\label{sec:implementation}


\subsection{Extracting and comparing features}
\label{sec:implementation-extracting}
Our implementation of BinPro has three major components.  The first component extracts functions and features from source code, while the second component extracts functions and features from binaries.  Finally, a third component performs function matching and labeling of the two sets of extracted functions and features from binary and source code.  The source code feature extraction component of BinPro is implemented by extending the ROSE compiler framework~\cite{ROSE-compiler} with 1907 Lines of Code (LOC). The source code extraction feature processes each source file individually and outputs each function as a description in Graphviz DOT format~\cite{graphviz}. The functions are then linked together along with the other source features in Table~\ref{table:featureSet} using a set of Python scripts (730 LOC) using PyDot~\cite{pydot} and Networkx~\cite{networks} libraries.



All binary features are extracted using IDA Pro~\cite{ida} and a Python script. To extract the number of arguments, we wrote a Python script that uses IDA Pro's API to extract the necessary binary features to compute the number of arguments.  On the x86-64 processor architecture, there are 6 dedicated registers used for passing arguments in a function call (\texttt{rdi, rsi, rdx, rcx, r8, r9})~\cite{x86_arg_reg}. However, compilers may optimize away definitions and uses of these registers, causing BinPro to only infer the number of arguments in the binary correctly 64\% of the time.  As a result, we found that machine learning tended to place a low weight factor on this feature, meaning that it only came into play when all the other features between two functions were very similar. 


Function matching is implemented in 12,110 lines of JAVA code. We have based our implementation on the Graph Matching Toolkit framework by Riesen and Bunke~\cite{bipartite-matching}, which provides an implementation of the Hungarian algorithm. 

\begin{table}[tb]
	\caption{Weight coefficients used in function matching.}
	\begin{center}
		\begin{tabular}{|l|r|}
			\hline
			\multicolumn{1}{|c|}{\textbf{Feature Name}} & 
			\multicolumn{1}{|c|}{\textbf{Weight}} \\ 
			\hline
			String constants & 1.469\\ \hline
			Integer constants & 0.6315\\ \hline
			Library calls & 0.2828\\ \hline
			FCG callers & 2.9293\\ \hline
			FCG callees & 2.9293\\ \hline
			\# of function args & 0.9296\\ \hline
			CFG branches &  0.0002 \\ \hline
		\end{tabular}
	\end{center}
	\label{table:weightsFCG}
	\vspace{-20pt}
\end{table}%

\subsection{Application of machine learning}
\label{sec:machineLearningImplementation}
We train machine learning classifiers for inlining prediction during feature extraction and to compute the optimal weights for edge weight calculation during the matching phase.  The likelihood of inlining and the feature weight coefficients is dependent on how the compiler applies optimizations, which depends on the optimization level used by the compiler.  For instance, there are four standard optimization levels in the GCC compiler.  GCC flag \texttt{-O0} will turn off all optimizations, whereas \texttt{-O1} will turn on 31 different optimizations. The flag \texttt{-O2} will turn on another 26 optimizations and \texttt{-O3} an additional 9.  BinPro is trained with a mix of the \texttt{O2} and \texttt{O3} optimizations levels on GCC, which are the most commonly used optimization levels for production code.

To train the inlining prediction classifier, we perform a 5-fold cross evaluation across our 10 applications. The applications in the training set do not have to have any relation to the application for which BinPro is eventually used to find the source code provenance.  Each application is compiled twice, once with \texttt{-O2} and again with \texttt{-O3} compiler optimization levels.  We use debugging symbols within the compiled binaries to divide the functions into an inlined set and a non-inlined set.  For each application, we then selected the smaller of the two sets (usually the set that was inlined) and randomly select an equal size set of functions from the other set, making a set of functions that is 50\% inlined and 50\% not-inlined, so that the training is not biased on either set.  
For each function, we extract a feature vector containing the predictive features for the function.  We then aggregate these functions and feature vectors across all applications to train our inlining predictor.  


The feature weight coefficients for function matching are computed using a similar procedure.  Again, we use a corpus of applications and compile each application using GCC, but only at the \texttt{O2} optimization level so that only functions that are very likely to be inlined get inlined.  We then exclude all the inlined functions and using the ground truth for which source code function matches which binary code function, we create a set of function pairs for each application composed of 50\% correct matches and 50\% incorrect matches.  Inlined functions are excluded because we cannot always combine the callee and caller features correctly, and these errors pollute the training set.  We then train the weights so that the SVM classifier is able to maximally classify these two sets correctly across all applications in the training set.  We perform a 5-fold cross evaluation across our 10 applications and tabulate the average computed weights from this 5-fold evaluation in Table~\ref{table:weightsFCG}, which shows that FCG callee and callers, as well as string constants are the main features used for matching as they have the heaviest weights. String constants are present in 55\% of functions and combined with their high discriminating power (as indicated by the high weight coefficient) they greatly contribute to BinPro's effectiveness.   The number of function arguments has some discriminating power, but is weighted lower because compilers may optimize away the uses and definitions of function arguments, causing BinPro to incorrectly extract the number of arguments from binaries 36\% of the time.  While still useful, integer constants and library calls have less discriminating power because they both tend to be fairly constant in many functions.

\myfigwide{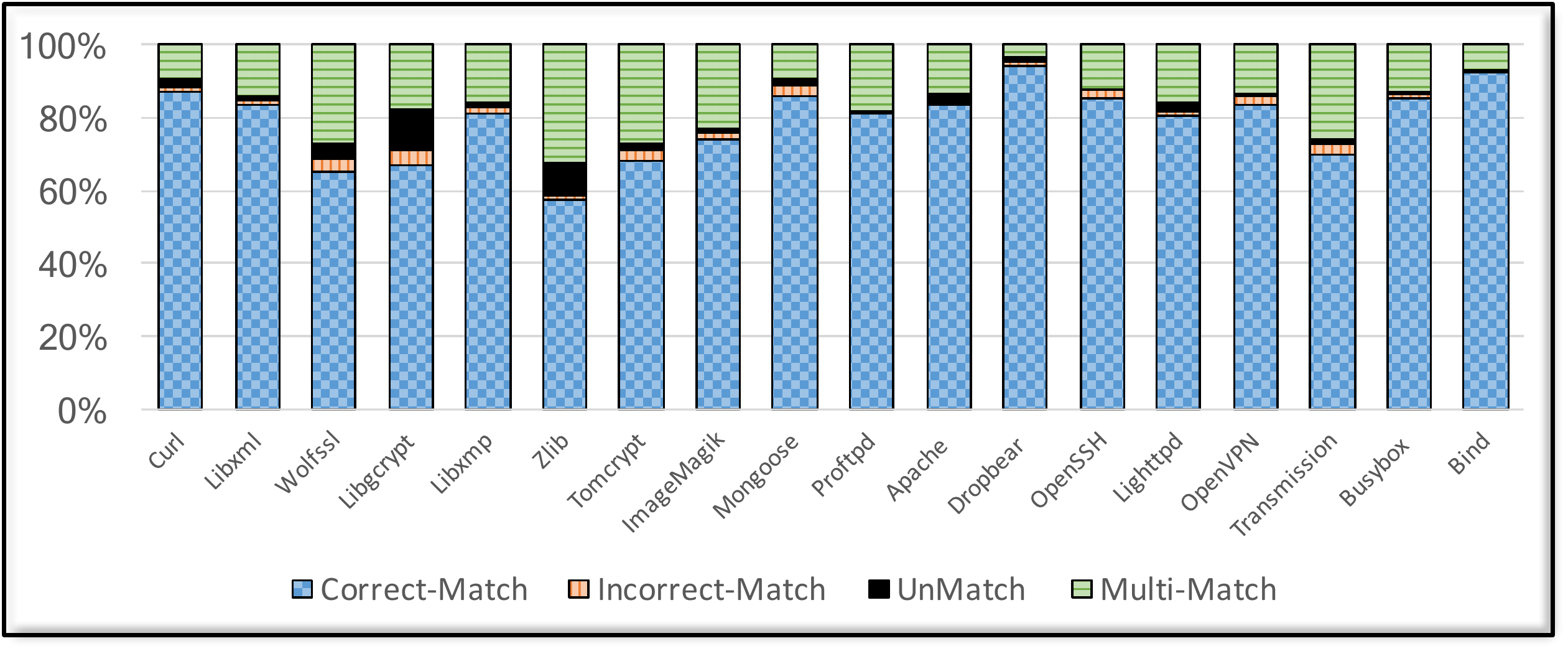}{BinPro matching accuracy on binaries and matching source code.}{fig:experiment1}{1.0}{-10pt}

As mentioned earlier, we originally thought BinPro could also use CFG features, such as the number of conditional branches in a function, as a feature for matching functions.  However, after training the classifier, we find that it assigns very little weight to the feature because the CFGs of functions change dramatically during compilation, and thus do not provide a good measure of similarity if standard compiler optimizations are applied.  In practice, using this noisy feature even with a low weight leads to more false matches, so BinPro does not use CFG features in its matching algorithm.

\section{Evaluation}
\label{sec:evaluation}

\myfigwide{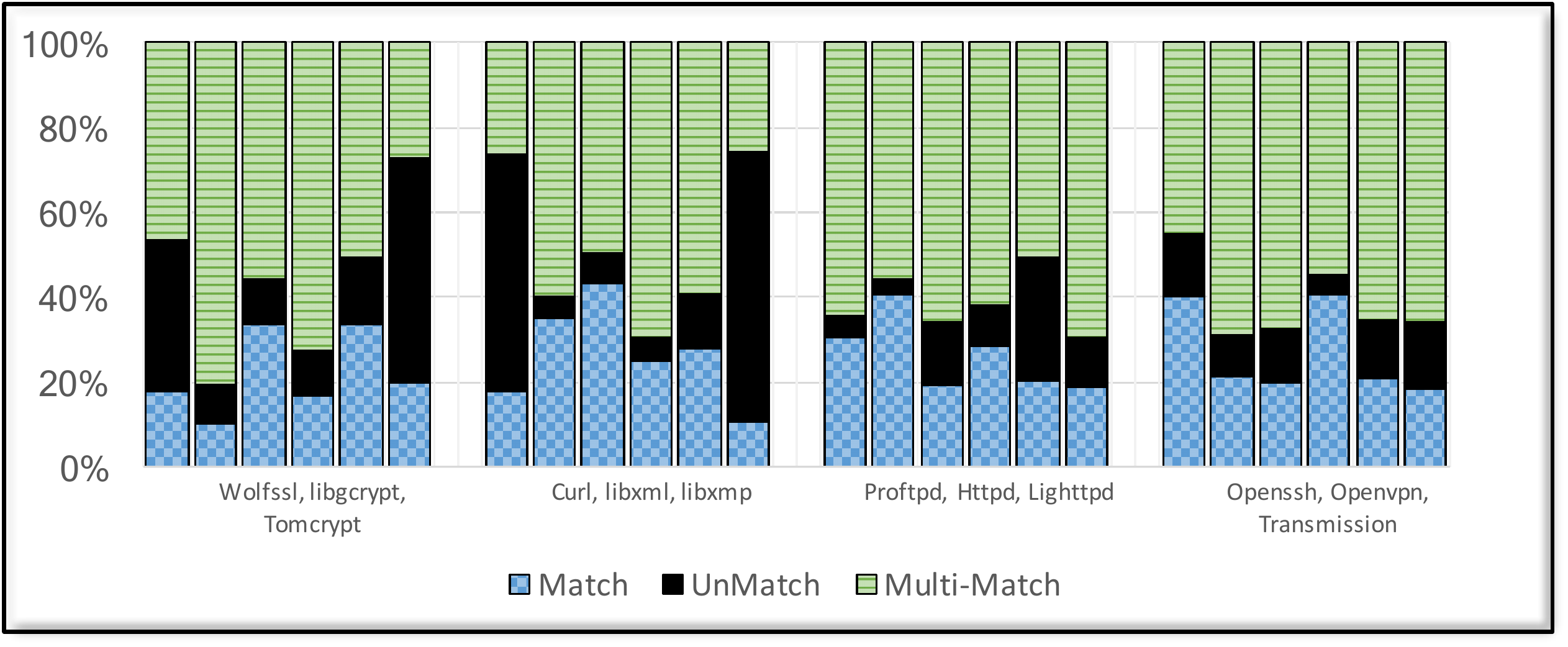}{BinPro's matching accuracy on four sets of three similar applications. Each bar represents BinPro's matching results on non-matching binary and source code.}{fig:experiment3}{1.0}{-10pt}

To see how effective BinPro is at computing similarity between a binary and source code, we evaluate BinPro on the following aspects:
\begin{itemize}
	\item {\bf Matching code:} How does BinPro score the similarity of a binary and its matching source code, i.e. the source code for the particular binary?
	\item {\bf Non-matching code:} How does BinPro score the similarity of a binary and its non-matching source code, i.e. the source code for a different binary?
	\item {\bf Compiler sensitivity:} How does BinPro's similarity score vary with binaries compiled with different compilers and different compiler optimization levels?
	\item {\bf Scalability:} How does BinPro's performance scale with code size and complexity?
	\item {\bf Inlining prediction:} How well does BinPro's inlining predictor work?
\end{itemize}

We evaluate BinPro on a set of 10 executable applications: Proftpd 1.3.4b, Busybox 1.23.2, Apache (HTTP) 2.4.12, Bind 9.10.2, Mongoose 2.1.0, OpenSSH 7.1p2, Dropbear 2015.71, OpenVPN 2.3.10, Transmission 2.84, Lighttpd 1.4.39,  and 8 libraries: libxml\footnote{\url{http://xmlsoft.org/}no prob}, Wolfssl 3.9.10, libgcrypt 1.7.2, libxmp 4.4.0\footnote{\url{http://xmp.sourceforge.net/}}, Zlib 1.2.8, Curl 7.50.3, Tomcrypt, ImageMagick.  For this study, we use virtual machines running Ubuntu 14.04 on Intel Core i7-2600 CPUs (4 cores @ 3.4 GHz) with 32 GB of memory.  

Our BinPro prototype currently supports only code that are fully compatible with GCC or the ICC Intel compiler because the ROSE framework used by BinPro does not work reliably with code written for other C/C++ compilers. BinPro also supports only x86-64 binary code because our IDAPro scripts are implemented for x86-64 binary code only.

\subsection{Matching code} 
\label{subsec:correctMatches}
We first evaluate BinPro's accuracy on a binary and the source code it was compiled from.  All applications in our dataset are open-source and we use their default build configuration  to compile binaries for this experiment. 
We then run BinPro with these binaries and their source code. Figure~\ref{fig:experiment1} gives the percentage of functions that BinPro labels as match, un-match and multi-match for each application.  All results are generated using a 5-fold cross-evaluation on the training sets used to build our machine learning models so that the applications used for training are never in the set of applications used for evaluation.  BinPro produces similarity scores between 59\% and 96\%, with an average of 81\%.   The reason BinPro produces low similarity score of 59\% for Zlib library is because Zlib is a very small application, consists of only 127 functions out of which 47 are library functions. A lot of these functions are very simple, with little-to-none features, causing BinPro to not find  unique matches to source code functions and instead, it finds large percentage of multi-matched functions.

\myfig{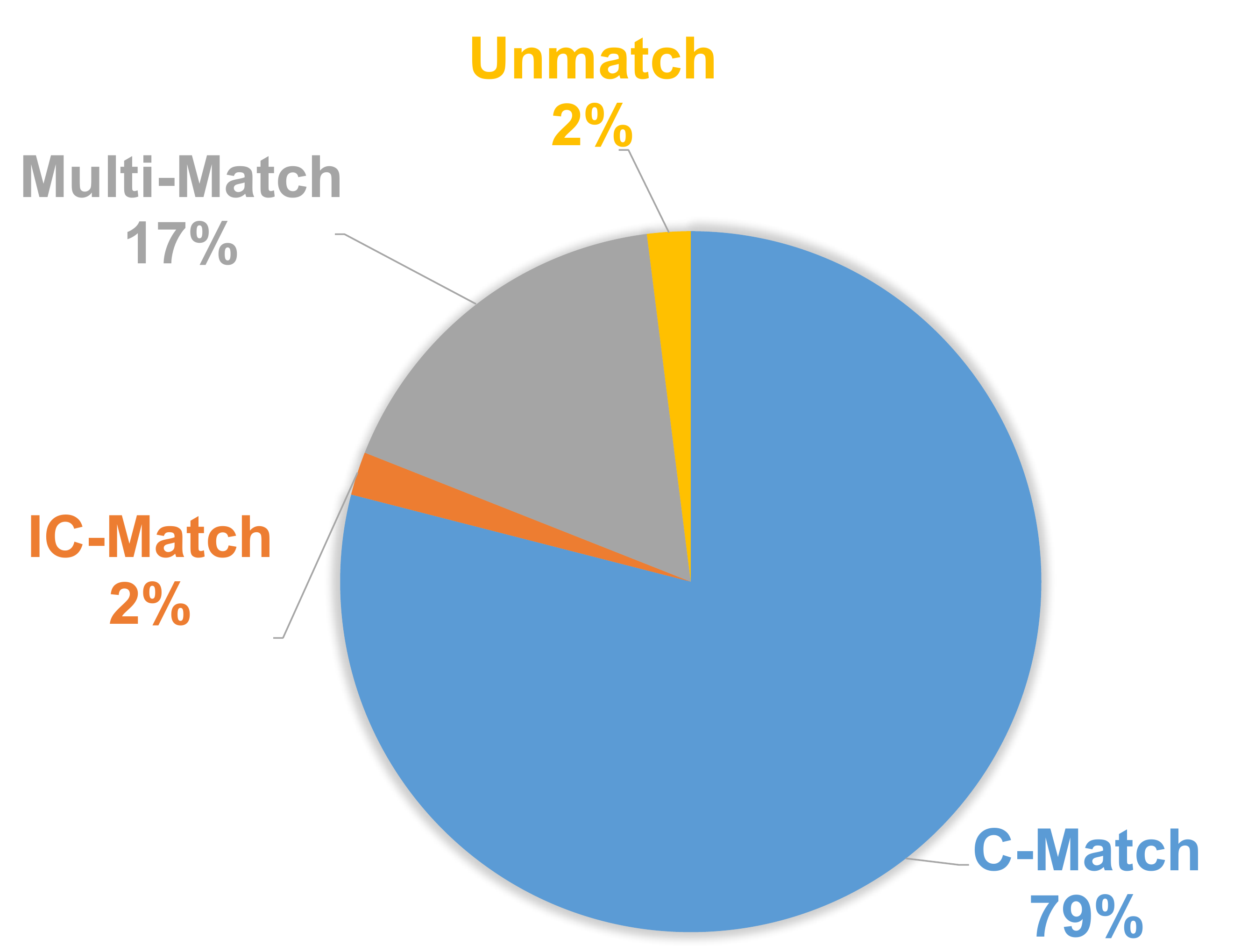}{Breakdown of BinPro's matching results.}{fig:pie}{1.0}{-10pt}

By generating debug symbols, we obtain the ground truth mapping between binary code functions and their corresponding source code functions.  Using this information, we compute how often function matching correctly assigns a binary code function to the correct source code function.  The number of function-pairs that BinPro labels as matched that are actually matched to the wrong functions in the application is also given in Figure~\ref{fig:experiment1}.  Figure~\ref{fig:pie} gives a breakdown of BinPro's average results across the applications.  ``C-'' means that BinPro correctly matched the binary code function while ``IC-'' means that BinPro matched the binary code function with the incorrect source code function.  Across the applications, BinPro correctly matches functions on average of 79.2\% of the time, whereas it incorrectly matches 1.9\% of functions. Finally, 16.6\% of functions are marked as multi-matched and 1.9\% as un-matched.

Upon closer inspection we find that the incorrectly matched functions tend to be simple and have few matching features, making it easy for BinPro to confuse them.  When examining the multi-matched functions, we find that the correct match is inside the set of candidates, but that all candidates just have very similar matching features.  Unmatched functions are all simple functions that have no matching features, and whose callers are multi-matched so the iterations during function matching never map them to a source code function.  

\subsection{Non-matching code}
We evaluate how BinPro scores the similarity of a binary and its non-matching source code.  We created 4 sets of 3 applications with similar functionality.  The first set, Wolfssl, libgcrypt and Tomcrypt are all libraries of cryptographic functions; the second set, Curl, libxml and libxmp are all applications that involve parsing complex inputs; the third set, ProFTPd, Httpd and Lighttpd are all web servers; and the fourth set, OpenSSH, OpenVPN and Transmission are all network utilities.  We run BinPro on the binary of each application against the source code of the other 2 applications in each set to obtain 6 measurements for each set.  Figure~\ref{fig:experiment3} shows the result of this experiment.  When the binary is not derived from the source code, BinPro computes significantly lower similarity scores.  BinPro produces scores from 10\% to 43\% with an average of 25\% across all 24 comparisons.  Compared with the the lowest similarity score of 59\% in the correct match case above, this shows that BinPro has good discriminating ability to identify if a binary was compiled from a particular source code.

\myfigwide{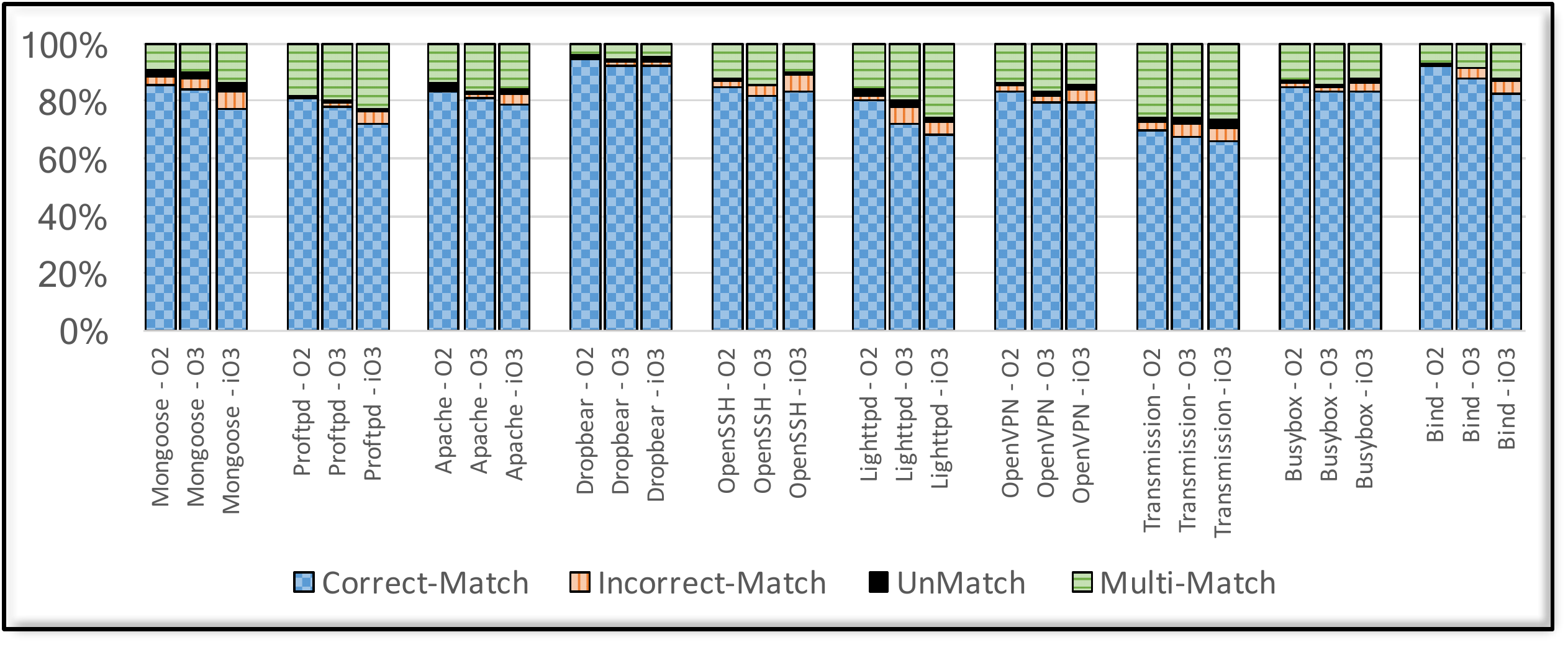}{BinPro's matching accuracy across different compilers and optimization levels. Each application is compiled with GCC -O2 and -O3 compilation flag, and Intel's ICC -O3 (iO3). }{fig:experiment2}{1.0}{-10pt}


\subsection{Compiler sensitivity} 
We now evaluate how sensitive BinPro's similarity scores are to different compilers and optimization levels.  We  select 10 applications from our dataset. Each application is compiled into three binaries, one compiled with GCC 4.8.2 using optimization level \texttt{O2}, one also with GCC using optimization level \texttt{O3}, and one compiled with ICC Intel compiler 15.0.3 using the highest optimization level \texttt{O3}.  We believe compilers aggressively perform source code optimization with level \texttt{O3} than \texttt{O2}. As a result, optimization level \texttt{O2} with ICC was not included in the evaluation.  

We follow the same procedure as in Section~\ref{subsec:correctMatches} to run BinPro on the matching code for this set of applications and present the result in Figure~\ref{fig:experiment2}. The largest change in similarity score due to changing compilers and optimization levels is 9.4\%, which we note is still smaller than the smallest difference between similarity scores between a binary and the source code it was compiled from and a binary and a different source code.  The average change due to different compilation is 4.0\%.  

Higher levels of optimization lead to more incorrectly matched functions, where BinPro labels a pair as matched, but the functions do not actually match.  However, even in the most adversarial case, where BinPro is trained on one compiler, GCC, and evaluated on binaries compiled with a completely different compiler, ICC, the rate of incorrectly matched function pairs only increases by an average of 2.5\%.  


\begin{table}[tb]
	\caption{BinPro execution time (in minutes) compared to application LOC (in thousands) and number of functions. }
	\begin{center}
		\begin{tabular}{|l|c|c|c|}
			\hline
			\multicolumn{1}{|c|}{\textbf{Application}} & 
			\multicolumn{1}{|c|}{\textbf{LOC}} & 
			\multicolumn{1}{|c|}{\textbf{\# of functions}} & 
			\multicolumn{1}{|c|}{\textbf{Time (min)}} \\ 
			\hline
			Zlib	& 25k & 127 & 0.02 \\ \hline
			Mongoose & 4k & 319 & 0.04\\ \hline
			Lighttpd & 101k & 458 & 0.5\\ \hline
			Dropbear 	 & 106k & 663 & 0.6\\ \hline
			Wolfssl	& 207k & 673 & 0.4 \\ \hline			
			libxmp	& 46k & 709 & 0.6\\ \hline
			Tomcrypt	& 102k & 900 & 1\\ \hline
			Curl	 & 237k & 1020 & 5\\ \hline
			libgcrypt	& 252k & 1303 & 5\\ \hline
			OpenVPN  & 119k & 1403 & 6\\ \hline
			Apache & 270k & 1270 & 3\\ \hline
			Proftpd & 300k & 1332 & 3\\ \hline
			OpenSSH & 143k & 1574 & 5\\ \hline
			Transmission	& 175k & 1574 & 5\\ \hline
			libxml	& 532k & 2472 & 9\\ \hline
			Busybox 	& 325k & 2664 & 75\\ \hline
			ImageMagick & 617k & 3268 & 110 \\ \hline
			Bind	& 619k & 4764 & 294\\ \hline
		\end{tabular}
	\end{center}
	\label{table:appsSize}
	\vspace{-10pt}
\end{table}

\subsection{Scalability}
We evaluate BinPro's execution time against the code size and the number of functions.  We find that BinPro execution time varies considerably, from several seconds for small applications with a few thousand lines of code and hundreds of functions, to almost 5 hours for Bind, which has more than half-a-million lines of code and almost 5 thousand functions.  BinPro's execution time is roughly proportional to the number of functions as opposed to code size.  We believe this is due to BinPro's algorithm, which extracts features and performs matching at a function granularity.  Details on application size and time it takes for BinPro to perform its analysis is given in Table~\ref{table:appsSize}. Upon closer profiling of BinPro, we see that on average 85\% of the execution of BinPro is used on loading the function call graphs (FCG) into memory.  Our virtual machines have a maximum of 32GB of memory and as a result larger applications that consume a large amount of memory are affected most from this restriction, as demonstrated by the rapid increase in execution time for large applications.  

\subsection{Inlining prediction}
Finally, we evaluate the inlining predictor in isolation to see how well it predicts when inline optimization will occur.  We performed a 5-fold cross evaluation across the same 10 applications as before. The predictor is trained on GCC as described in Section~\ref{sec:machineLearningImplementation} and then evaluated on the Intel ICC compiler.   The inlining prediction correctly predicts inlining on 73\% of inlined functions and incorrectly predicts inlining on 17\% of non-inlined functions.  This confirms that the accuracy of the inline predictor across different compilers is one of the reasons that BinPro is able to achieve high accuracy even across different compilers.

\section{Conclusion}
\label{sec:conclusion}

BinPro measures similarity between a binary and a source code, and with this measurement, is able to determine if a binary was compiled from a source code it is compared with.  When evaluated across 18 executable applications and libraries, BinPro computes an average similarity of 81\% when binary and source code match, and only 25\% when they don't match.  Moreover, the highest similarity for a non-matching pair of binary and source code is 43\% while the lowest similarity for a matching pair is 59\% indicating that BinPro can accurately tell if a binary was compiled from a source code with no incorrect results.

BinPro's most interesting result is that it is able to correctly match functions in binaries with their source code counterparts 79\% of the time on average, regardless of the compiler or optimization level used.  BinPro achieves this by using machine learning to predict when compiler optimizations might be applied and uses a set of matching code features combined with bipartite matching to find the best match between binary code and source code functions.  By repeatedly performing the matching, BinPro is able iteratively incorporate function call graph callers and callees into the matching process.  The careful selection of matching features, use of machine learning, use of bipartite matching and iterative matching all contribute to BinPro's accuracy.

\bibliographystyle{ACM-Reference-Format}
\bibliography{bibfile}

\end{document}